\documentclass[fleqn,preprint,showpacs,preprintnumbers,amsmath,amssymb]{revtex4}
\usepackage{graphicx}
\usepackage{dcolumn}
\usepackage{bm}

\begin{document}

\title{Propagation of arbitrary-amplitude ion waves in relativistically degenerate electron-ion plasmas}
\author{M. Akbari-Moghanjoughi}
\affiliation{ Azarbaijan University of
Tarbiat Moallem, Faculty of Sciences,
Department of Physics, 51745-406, Tabriz, Iran}

\date{\today}
\begin{abstract}
We employ the Sagdeev pseudo-potential method to investigate the propagation of nonlinear ion waves in a relativistically degenerate electron-ion plasmas. The matching criteria for existence of such nonlinear excitations are numerically investigated in terms of the relativity measure (relativistic degeneracy parameter) of electrons and the allowed Mach-number range for propagation of such waves is evaluated. It is shown that the electron relativistic degeneracy parameter has significant effects on nonlinear wave dynamics in superdense degenerate plasmas such as that encountered in white dwarfs and the cores of massive planets.\\ 
(\textbf{Accepted for publication in Astrophysics \& Space Science})
\end{abstract}


\maketitle

\section{Introduction}

Recently, investigation of the plasma under extreme conditions, such as high temperature and pressure, has attracted outstanding attention because of numerous applications in laboratory \cite{haug, son1, son2, greaves} as well as astrophysical environments \cite{shapiro, rees, miller, goldr, michel}. Under such conditions the quantum and relativistic effects are unavoidable and one must take into account the relativistic degeneracy pressure which arises due to the Pauli-blocking mechanism. The quantum effects become more important when the inter-fermion distances are much lower than the characteristic de Broglie thermal wavelength, $\hbar/(2\pi m k_B T)^{1/2}$ \cite{bonitz, shukla}. Chandrasekhar \cite{chandra1}, using the Fermi-Dirac statistics for electrons, has developed a rigorous mathematical criteria under which a white dwarf can be considered as completely degenerate, i.e. zero-temperature Fermi-gas. The complete degeneracy may seem a simplifying assumption, however, it hold for many dense astrophysical entities such as white dwarfs \cite{chandra2, misner}. For a completely degenerate plasma one can assume well defined Fermi quantities such as that for typical metals.

The electron degeneracy in massive white dwarfs which opposes the catastrophic gravitational inward pressure can give rise to relativistic degeneracy at some condition. Many studies in the framework of quantum hydrodynamics dealing with astrophysical models consider only the non-relativistic degeneracy cases \cite{sabry, misra1, misra2, chatterjee, akbari1, abdelsalam}. However, in a typical white dwarf where the electron density can be as high as $10^{28}cm^{-3}$ the degeneracy can be relativistic and both quantum and relativity considerations should be taken into account. Semiclassical approach shows that the dynamics of nonlinear ion-acoustic propagations in an electron-positron-ion can be much different under relativistic degeneracy pressure from those under non-relativistic one \cite{akbari2, akbari3}. The presented investigation based on the quantum hydrodynamics model does not cover all range of the relativity parameter for plasma under consideration and are relevant only to either non-relativistic or ultra-relativistic cases.

The aim of current investigation is to study the propagation of nonlinear excitations, namely, ion solitary waves (ISWs) and ion periodic waves (IPWs) with arbitrary amplitudes in relativistically degenerate zero-temperature electron-ion plasma. Although, it has been confirmed that the model of zero-temperature Fermi-gas is quite appropriate for most white dwarfs \cite{chandra1}, however, the full determination of large-amplitude propagation of ISWs and IPWs in such plasma with extreme temperatures requires the finite-temperature Fermi-Dirac model which accounts for the electron temperature that may be our future work. The presentation of the article is as follows. The basic normalized plasma equations are introduced in section \ref{equations}. Nonlinear arbitrary-amplitude solutions are derived in section \ref{Sagdeev}. Numerical analysis and discussions is given in section \ref{discussion} and final remarks are presented in section \ref{conclusion}.

\section{Basic Hydrodynamics Equations}\label{equations}

We consider a completely degenerate dense plasma consisting of relativistic cold-electrons and dynamic cold-ions. It is noted that in degenerate plasmas the rate of electron-ion collisions are limited due to Pauli-blocking mechanism which allows only degenerate particles with energies limited to a narrow range around the Fermi-energies to interact, hence, the plasma may be considered almost collision-less. We use the conventional fluid hydrodynamics equations, which in dimensional form can be written as
\begin{equation}\label{dim}
\begin{array}{l}
\frac{{\partial {n_i}}}{{\partial t}} + \frac{{\partial {n_i}{v_i}}}{{\partial x}} = 0, \\
\frac{{\partial {n_e}}}{{\partial t}} + \frac{{\partial {n_e}{v_e}}}{{\partial x}} = 0, \\
\frac{{\partial {v_i}}}{{\partial t}} + {v_i}\frac{{\partial {v_i}}}{{\partial x}} =  - \frac{1}{{{m_i}}}\frac{{\partial \phi }}{{\partial x}} - \frac{{\gamma {k_B}{T_i}}}{{{m_i n_{i0}}}}\left(\frac{n_i}{n_{i0}}\right)^{\gamma  - 2}\frac{{\partial {n_i}}}{{\partial x}}, \\
\frac{{{m_e}}}{{{m_i}}}\left( {\frac{{\partial {v_e}}}{{\partial t}} + {v_e}\frac{{\partial {v_e}}}{{\partial x}}} \right) = \frac{1}{{{m_i}}}\frac{{\partial \phi }}{{\partial x}} - \frac{1}{{{m_i}{n_e}}}\frac{{\partial {P_e}}}{{\partial x}} + \frac{{{\hbar ^2}}}{{2{m_e}{m_i}}}\frac{\partial }{{\partial x}}\left( {\frac{1}{{\sqrt {{n_e}} }}\frac{{{\partial ^2}\sqrt {{n_e}} }}{{\partial {x^2}}}} \right), \\ \frac{{{\partial ^2}\phi }}{{\partial {x^2}}} = 4\pi e({n_e} - {n_i}), \\
\end{array}
\end{equation}
In order to obtain the normalized set of equations, following scalings can be used
\begin{equation}\label{T}
x \to \frac{{{c_s}}}{{{\omega _{pi}}}}\bar x,t \to \frac{{\bar t}}{{{\omega _{pi}}}},n \to \bar n{n_0},u \to \bar u{c_s},\phi  \to \bar \phi \frac{{{m_e}{c^2}}}{e}.
\end{equation}
where, ${\omega _{pi}} = \sqrt {4\pi {e^2}{n_{e0}}/{m_i}}$ and ${c_{s}} = \sqrt {{m_e} {c^2}/{m_i }}$ are characteristic plasma frequency and sound-speed, respectively. Also, $n_{e0}$ denotes the electron equilibrium density. Furthermore, the electron degeneracy pressure in fully degenerate configuration can be expressed in the following form \cite{chandra1}
\begin{equation}
{P_e} = \frac{{\pi m_\alpha ^4{c^5}}}{{3{h^3}}}\left\{ {{R}\left( {2{R^2} - 3} \right)\sqrt {1 + {R^2}}  + 3\ln \left[ {R + \sqrt {1 + {R^2}} } \right]} \right\},
\end{equation}
in which $R=p_{Fe}/m_e c=(n_e/n_0)^{1/3}$ (${n_0} = \frac{{8\pi m_\alpha^3{c^3}}}{{3{h^3}}}\simeq 5.9 \times 10^{29} cm^{-3}$), where, $p_{Fe}$ is the electron Fermi relativistic-momentum. It is noted that in the limits of very small and very large values of the relativity parameter, $R$, we obtain (e.g. see Ref. \cite{chandra1})
\begin{equation}\label{limits}
P_e = \left\{ {\begin{array}{*{20}{c}}
{\frac{1}{{20}}{{\left( {\frac{3}{\pi }} \right)}^{\frac{2}{3}}}\frac{{{h^2}}}{{{m_e}}}n_e^{\frac{5}{3}}\hspace{10mm}(R \to 0)}  \\
{\frac{1}{{8}}{{\left( {\frac{3}{\pi }} \right)}^{\frac{1}{3}}}hcn_e^{\frac{4}{3}}\hspace{10mm}(R \to \infty )}  \\
\end{array}} \right\}.
\end{equation}
Therefore, the relativistic normalized hydrodynamics set of plasma equations including the electron tunneling term, considering the fact that $1/{n_e}(\partial {P_e}/\partial x) = \partial \sqrt {1 + {R^2}} /\partial x$, may be written as
\begin{equation}\label{normal}
\begin{array}{l}
\frac{{\partial {n_i}}}{{\partial t}} + \frac{{\partial {n_i}{v_i}}}{{\partial x}} = 0, \\
\frac{{\partial {n_e}}}{{\partial t}} + \frac{{\partial {n_e}{v_e}}}{{\partial x}} = 0, \\
\frac{{\partial {v_i}}}{{\partial t}} + {v_i}\frac{{\partial {v_i}}}{{\partial x}} =  - \frac{{\partial \phi }}{{\partial x}} - \left( {\frac{{\gamma {k_B}{T_i}}}{{n_{i0}{m_e}{c^2}}}} \right)\left(\frac{n_i}{n_{i0}}\right)^{\gamma  - 2}\frac{{\partial {n_i}}}{{\partial x}}, \\
\frac{{{m_e}}}{{{m_i}}}\left( {\frac{{\partial {v_e}}}{{\partial t}} + {v_e}\frac{{\partial {v_e}}}{{\partial x}}} \right) = \frac{{\partial \phi }}{{\partial x}} - \frac{{\partial \sqrt {1 + {R_0^{2} {n_e}^{2/3}}} }}{{\partial x}} + \frac{{{H_r^{2}}}}{2}\left( {\frac{{{m_e}}}{{{m_i}}}} \right)\frac{\partial }{{\partial x}}\left( {\frac{1}{{\sqrt {{n_e}} }}\frac{{{\partial ^2}\sqrt {{n_e}} }}{{\partial {x^2}}}} \right), \\
\frac{{{\partial ^2}\phi }}{{\partial {x^2}}} = {n_e} - {n_i}, \\
\end{array}
\end{equation}
where, we have used $R=R_0 n_e^{1/3}$ and $R_0=(n_{e0}/n_0)^{1/3}$ is a measure of the relativistic effects (i.e. relativistic degeneracy parameter) and $H_r^{2}/2=\hbar^2 \omega_{pi}^2/(m_e c^2)^2$, is the relativistic quantum diffraction parameter which is related to the relativistic degeneracy parameter via the relation $(m_e/m_i)H_r^{2}/2\simeq 4.6\times 10^{-10} R_0^{3}$ (${\omega _{pi}} = \sqrt {4\pi {e^2}{n_0}{R_0^{3}}/{m_i}}$). It should be noted that the parameter $R_0$ is related also to the mass-density (of white dwarf, for instance) through the relation $\rho\simeq 2m_p n_{e0}$ or $\rho(gr/cm^{3})=\rho_0 R_0^{3}$ with $\rho_0(gr/cm^{3})\simeq 1.97\times 10^6$, where, $m_p$ is the proton mass. The density $\rho_0$ is exactly in the range of a mass-density of a typical white dwarf. The density of typical white dwarfs can be in the range $10^{5}<\rho(gr/cm^{3})<10^{9}$, which results in values of $0.37<R_0<8$ for the relativity parameter. S. Chandrasekhar, in ground of Fermi-Dirac statistics, has proven that for a white dwarf with a mass density $\rho$, the electron degeneracy pressure turns from $P_e\propto \rho^{5/3}$ dependence for normal degeneracy ($R_0 \ll 1$) to $P_e\propto \rho^{4/3}$ dependence for ultra-relativistic electron degeneracy ($R_0 \gg 1$) \cite{chandra2}. For the highest value of $R_0=8$ (as for the typical white-dwarf), we get the negligible value of $(m_e/m_i)H_r^{2}/2\simeq 10^{-13}$, which validates the applicability of current model to white dwarfs or even neutron stars. In the forthcoming analysis higher values of the relativistic degeneracy parameter has also been considered, since, for neutron stars the electron densities of order $n_e\simeq 10^{30}(cm^{-3})$ is possible. More recently, A. A. Mamun and P. K. Shukla \cite{mamun} have considered the nonlinear solitary excitations and double-layers in dusty degenerate dense plasmas in ultra-relativistic limit, however, as it is noted from Eqs. (\ref{limits}), their corresponding limit is only a special case, while, the present model includes the whole range of relativistic degeneracy parameter, $R_0$. Now, neglecting the term containing the ratio $m_e/m_i$ and assuming that $k_B T_i \ll m_e c^2$, we arrive at the following simplified set of basic normalized equations
\begin{equation}\label{comp}
\begin{array}{l}
\frac{{\partial {n_i}}}{{\partial t}} + \frac{{\partial {n_i}{v_i}}}{{\partial x}} = 0, \\
\frac{{\partial {n_e}}}{{\partial t}} + \frac{{\partial {n_e}{v_e}}}{{\partial x}} = 0, \\
\frac{{\partial {v_i}}}{{\partial t}} + {v_i}\frac{{\partial {v_i}}}{{\partial x}} =  - \frac{{\partial \phi }}{{\partial x}}, \\
\frac{{\partial \phi }}{{\partial x}} - \frac{{\partial \sqrt {1 + {R_0^{2} {n_e}^{2/3}}} }}{{\partial x}}=0, \\
\frac{{{\partial ^2}\phi }}{{\partial {x^2}}} = {n_e} - {n_i}. \\
\end{array}
\end{equation}
Solving the basic equations (Eqs. (\ref{comp})) for the electron and ion number densities in terms of electrostatic potential, at the quasi-neutrality condition, and employing the appropriate boundary requirement ($\mathop {\lim }\limits_{{v_{e,i}} \to 0} {n_{e,i}} = 1$ and $\mathop {\lim }\limits_{{v_{e,i}} \to 0} \phi  = 0$), results in the following energy-density relations
\begin{equation}\label{phis}
\begin{array}{l}
{n_i} = \frac{1}{{\sqrt {1 - \frac{{2\phi }}{{{M^2}}}} }}, \\
{n_e} = R_0^{-3}{\left[ {R_0^{2} + \phi \left( {2\sqrt {1 + R_0^{2}}  + \phi } \right)} \right]^{3/2}}. \\
\end{array}
\end{equation}

These equations are somewhat different from the standard energy-density relations for non-relativistic semiclassical Thomas-Fermi model \cite{akbari2, akbari3, abdelsalam}, as expected. Note that, unlike for Thomas-Fermi model for plasma, in Eqs. (\ref{phis}) there is no explicit contribution for electron (Fermi) temperatures, since, there is a hidden dependence with the electron equilibrium quantum number-densities. It is worth mentioning that, there is a widespread misconceptions about the independence of the particles equilibrium number-density and their corresponding Fermi-temperatures in many recent quantum plasma literatures \cite{comment}, which is avoided in current model. The state of a completely degenerate quantum plasma is determined through only one of the mentioned quantities since the Fermi-quantities are fixed via the particle equilibrium density. In the proceeding section we aim at finding the appropriate pseudo-potential which describes the propagation of nonlinear ion-acoustic waves in the plasma described by Eq. (\ref{comp}).

\section{Arbitrary-amplitude Nonlinear Propagations}\label{Sagdeev}

To obtain the appropriate Sagdeev pseudo-potentials which describes the possibility of ISWs as well as IPWs in a relativistically dense electron-ion plasma, we may reduce to stationary frame by moving into the new coordinate $\xi=x-Mt$, where, $M$ is the Mach number being a measure of the wave-speed relative to that of plasma sound, $c_s$. Substituting Eqs. (\ref{phis}) into Poisson's equation, multiplying by $\frac{{d\phi }}{{d\xi }}$ and integrating once with suitable boundary conditions, mentioned in the previous section, we derive the well-known energy integral
\begin{equation}\label{energy}
\frac{1}{2}{\left( {\frac{{d\phi }}{{d\xi }}} \right)^2} + U(\phi ) = 0,
\end{equation}
where, the desired Sagdeev pseudo-potentials $U(\phi)$ read as
\begin{equation}\label{S1}
\begin{array}{l}
U(\phi ) = \frac{1}{8}\left\{ {8{M^2}R_0^{3} + {R_0}\left( {2R_0^{2} - 3} \right)\sqrt {1 + R_0^{2}}  - 8{M^2}R_0^{3}\sqrt {1 - \frac{{2\phi }}{{{M^2}}}} } \right. \\
- \left[ {2{R_0^{2}} - 3 + 2\phi \left( {2\sqrt {1 + {R_0^{2}}}  + \phi } \right)} \right]\left( {\sqrt {1 + {R_0^{2}}}  + \phi } \right)\sqrt {R_0^{2} + \phi \left( {2\sqrt {1 + R_0^{2}}  + \phi } \right)}  \\
\left. { - 3\ln 2\left[ {\sqrt {1 + R_0^{2}}  + \phi  + \sqrt {R_0^{2} + \phi \left( {2\sqrt {1 + R_0^{2}}  + \phi } \right)} } \right] + 3{{\sinh }^{ - 1}}{R_0} + \ln 8} \right\}. \\
\end{array}
\end{equation}
Note that the values of $U(\phi)$ and ${\partial _\phi }U(\phi )$ both vanish at $\phi=0$, as required. From Eqs. (\ref{phis}) we may also derive the maximum values of $\phi$ as
\begin{equation}\label{phim1}
{\phi _{m1}} = \frac{{{M^2}}}{2},
\end{equation}
\begin{equation}\label{phim3}
{\phi _{m2}} =  \pm 1 - \sqrt {1 + R_0^{2}},
\end{equation}
The possibility of solitary waves then requires the following conditions to met, simultaneously
\begin{equation}\label{conditions}
{\left. {U(\phi)} \right|_{\phi = 0}} = {\left. {\frac{{dU(\phi)}}{{d\phi}}} \right|_{\phi = 0}} = 0,{\left. {\frac{{{d^2}U(\phi)}}{{d{\phi^2}}}} \right|_{\phi = 0}} < 0.
\end{equation}
It is further required that for at least one either maximum or minimum nonzero $\phi$-value, we have $U(\phi_{m})=0$, so that for every value of $\phi$ in the range ${\phi _m} > \phi  > 0$ or ${\phi _m} < \phi  < 0$, $U(\phi)$ is negative. In such a condition we can obtain a potential minimum which describes the possibility of a ISW propagation.

On the other hand, the existence of a nonlinear periodic wave requires that
\begin{equation}\label{conditions2}
{\left. {U(\phi)} \right|_{\phi = 0}} = {\left. {\frac{{dU(\phi)}}{{d\phi}}} \right|_{\phi = 0}} = 0,{\left. {\frac{{{d^2}U(\phi)}}{{d{\phi^2}}}} \right|_{\phi = 0}} > 0,
\end{equation}
It is also required that for at least one either maximum or minimum nonzero $\phi$-value, we have $U(\phi_{m})=0$, so that for every value of $\phi$ in the range ${\phi _m} > \phi  > 0$ or ${\phi _m} < \phi  < 0$, $U(\phi)$ is positive. In such a condition we can obtain a potential minimum which describes the possibility of a IPW propagation.

Evaluation of the second derivative of the Sagdeev pseudo-potential immediately yields a critical Mach-number value of
\begin{equation}\label{Mcr}
{M_{cr}} = \frac{{R_0{{(1 + R_0^{2})}^{ - 1/4}}}}{{\sqrt 3 }},
\end{equation}
below/above which IASW/IAPW may exist. This value can also be found considering the nature of these nonlinear waves whose amplitude tends to zero as the Mach-number $M$ tends to its critical value, $M_{cr}$ which may be predicted by expanding the pseudo-potential $U(\phi)$ to the third order in a Taylor series in $\delta \phi$ near $\phi =0$. However, since, the other conditions on the pseudo-potential requires analytic solutions, which is nontrivial to obtain in our case, in the proceeding section a numerical scheme should be used.

\section{Numerical Analysis and Discussion}\label{discussion}

Figure 1 shows the region of existence for ISWs/IPWs in yellow/blue (right/left region) in two different scales in $M$-$R_0$ plane. It is remarked that the allowed Mach-range is significantly affected by the relativity parameter, $R_0$. It is also observed that the IPWs posses relatively lower Mach-values compared to ISWs for all values of the relativistic degeneracy parameter, $R_0$. It is further observed that as the relativity parameter increases the Mach-range for both nonlinear excitation (ISWs and IPWs) increase and the Mach-number range tend to higher values. Figure 1(a) shows that supersonic solitary excitations occur beyond the values $R_0>1.3$. This value ($R_0>1.3$) corresponds to a mass-density of $\rho\simeq 4.3\times 10^{6}(gr/cm^{3})$ which is within the range of mass-density of typical white dwarfs ($10^{5}<\rho(gr/cm^{3})<10^{9}$). It should be noted, however, that the parameter $M=v_g/c_{s}$ is the relativistic Mach-number, since, the normalization parameter $c_{s}\simeq 8\times 10^{8} (cm/s)$ is comparable to the Fermi-speed of electrons in typical metals and is much higher than the ion-acoustic speed in the non-relativistic non-degenerate plasmas. It should be noted also that for all the values of the relativistic degeneracy parameter, $R_0$, there is a lower Mach-number limit for the propagation of both types of nonlinear excitations (ISWs and IPWs). On the other hand, Fig. 1(b) shows that both nonlinear excitations lie entirely in supersonic range for relativistic parameters $R_0>25$. This value ($R_0>25$) corresponds to a mass-density of $\rho\simeq 3\times 10^{10}(gr/cm^{3})$, which is larger than that for even a massive white dwarf.

The profiles of pseudo-potentials and their variation with respect to the relativity parameter and Mach-number are shown in Fig. 2 for both ISWs and IPWs. It is remarked that, the increase of the relativistic degeneracy parameter $R_0$ decreases/increases the potential depth and width for IASWs/IAPWs. Also, it is observed that the increase in the Mach-number increases/decreases the potential width and depth for ISWs/IPWs. On the other hand, Fig. 3(a) shows the solitary excitation profiles for different relativistic degeneracy parameter values for a fixed Mach-number. It is confirmed that the amplitude of ISW reduces with increase of $R_0$ (which is directly related to the mass-density), while it increases with increases in the value of the Mach-number for a fixed $R_0$ value (Fig. 1(b)). Therefore, it is confirmed that the narrower/wider the solitary wave the taller/smaller it is, which is a general nonlinear wave feature in agreement with the relation $\phi_0^{2}\Lambda=cst.$, where, $\phi_0$ and $\Lambda$ denote the amplitude and the width of nonlinear waves, respectively.

\section{Concluding Remarks}\label{conclusion}

The Sagdeev pseudo-potential approach was used to investigate the propagation of nonlinear ion waves in a superdense electron-ion plasma with relativistically degenerate electrons. The matching criteria of existence of such solitary excitations were numerically investigated in terms of the relativity measure of electrons and the allowed Mach-number range for propagation of such waves was evaluated numerically. It was shown that the relativistic degeneracy of electron has crucial effects on nonlinear wave dynamics in relativistically degenerate plasmas such as that found in white dwarfs, neutron stars and the cores of massive planets.

\newpage

\textbf{FIGURE CAPTIONS}

\bigskip

Figure-1

\bigskip

(Color online) The stability regions of arbitrary-amplitude IASWs/IAPWs is shown in $M$-$R_0$ plane for a relativistic degenerate electron-ion plasma. The lower region (yellow) correspond to the stability of ISWs and the upper (blue) region to IPWs.

\bigskip

Figure-2

\bigskip

(Color online) The variation of Sagdeev pseudo-potential with respect to different relativity parameter, $R_0$, and Mach-number, $M$ for both ISWs and IPWs.

\bigskip

Figure-3

\bigskip

(Color online) The profiles of solitary waves and the variation of their shape with respect to different relativity parameter, $R_0$, and Mach-number, $M$.

\newpage

\begin{figure}
\resizebox{1\columnwidth}{!}{\includegraphics{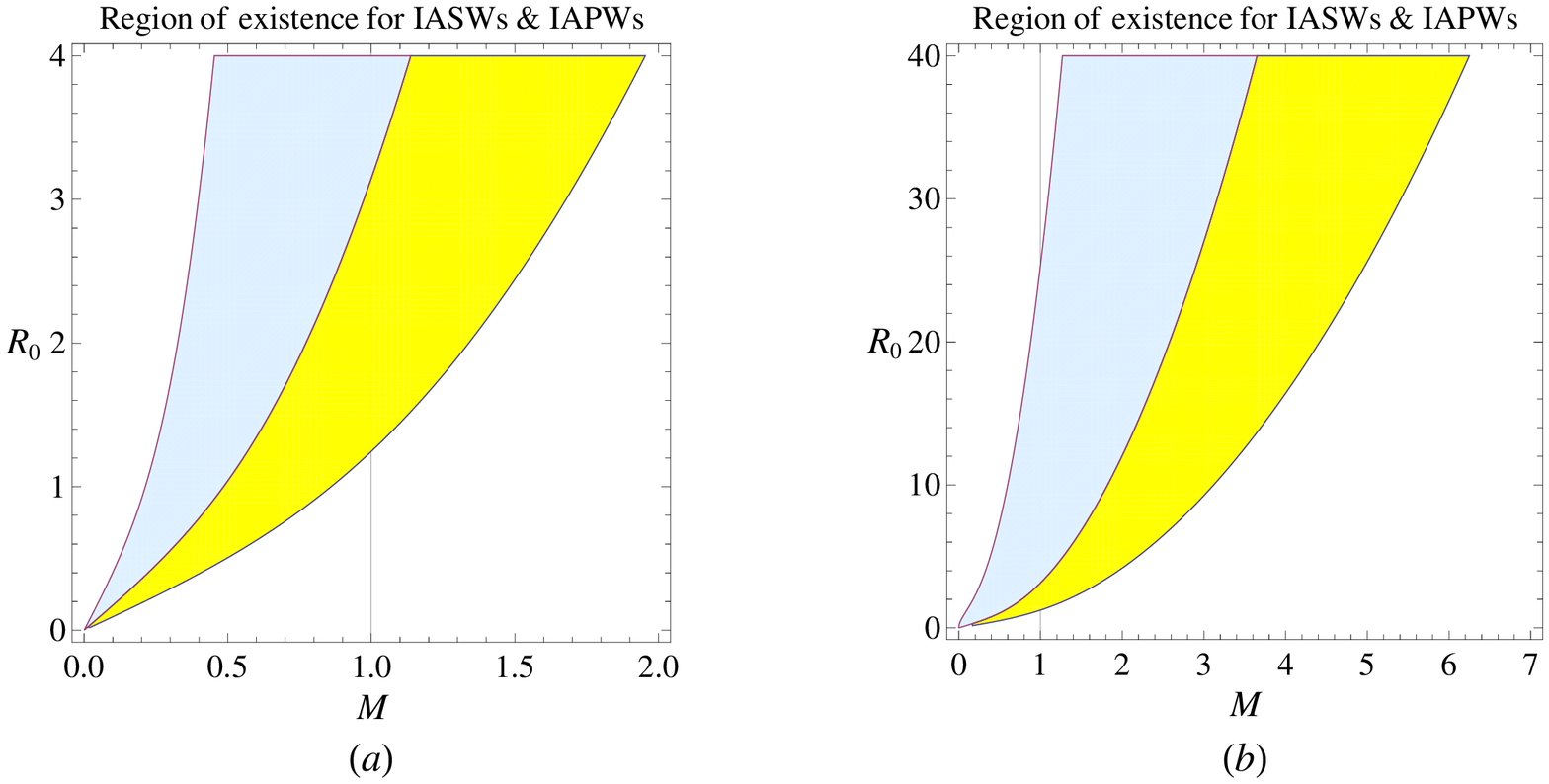}}
\caption{}
\label{fig:1}
\end{figure}

\newpage

\begin{figure}
\resizebox{1\columnwidth}{!}{\includegraphics{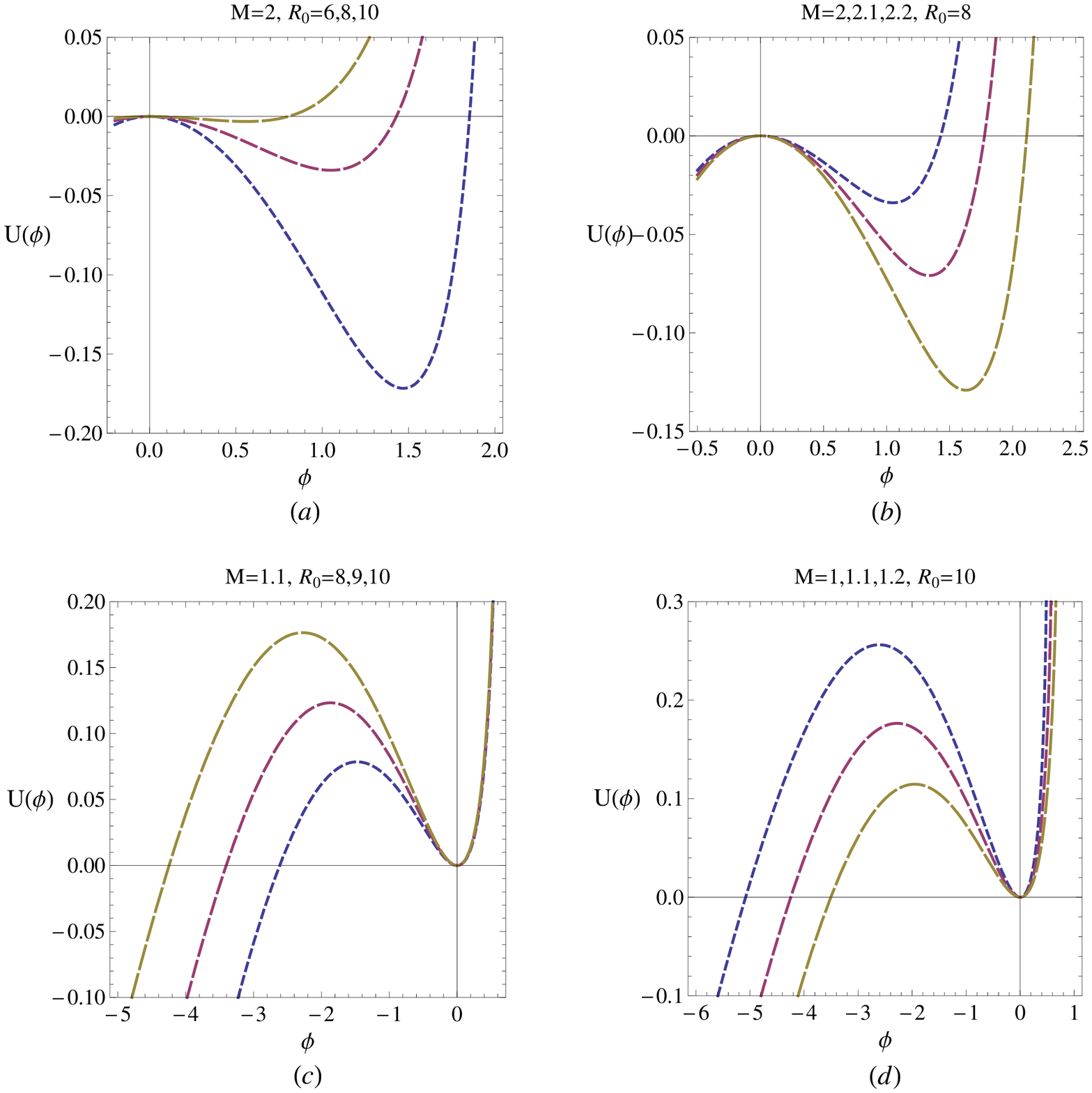}}
\caption{}
\label{fig:2}
\end{figure}

\newpage

\begin{figure}
\resizebox{1\columnwidth}{!}{\includegraphics{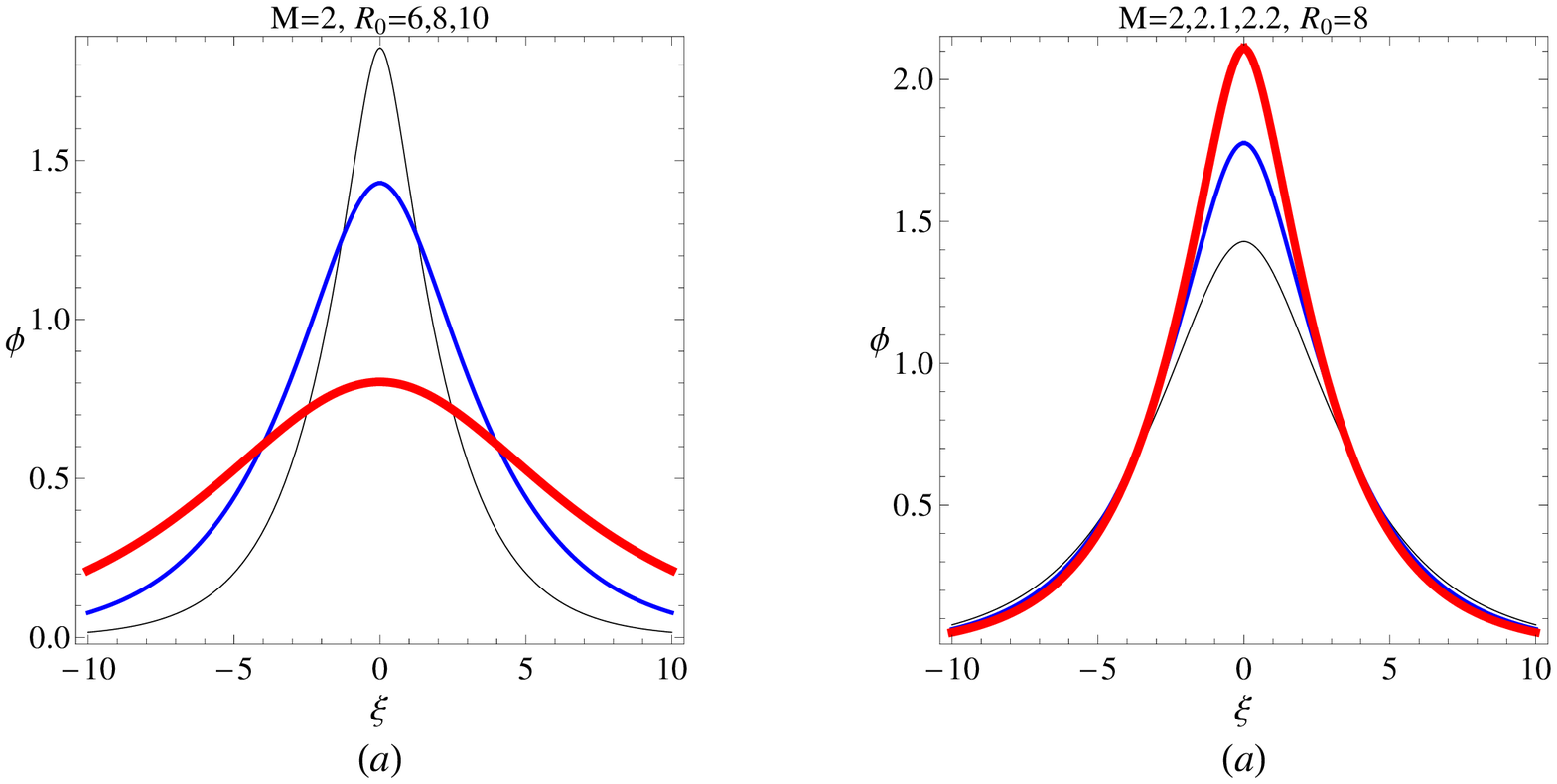}}
\caption{}
\label{fig:3}
\end{figure}

\end{document}